\documentclass[12pt,preprint]{aastex}
\usepackage{graphicx}

\begin{document}

\title{Molecular Line Observations of Carbon-Chain-Producing Regions L1495B and L1521B}
\author{Tomoya HIROTA, Hiroyuki MAEZAWA}
\affil {National Astronomical Observatory of Japan,}
\affil {Osawa 2-21-1, Mitaka, Tokyo 181-8588, Japan; 
tomoya.hirota@nao.ac.jp}
\and
\author{Satoshi YAMAMOTO}
\affil {Department of Physics and Research Center for the Early Universe,}
\affil {The University of Tokyo, Bunkyo-ku, Tokyo 113-0033, JAPAN }

\begin{abstract}

We present the first comprehensive study on physical and 
chemical properties of quiescent starless cores L1495B and L1521B, 
which are known to be rich in carbon-chain molecules like 
the cyanopolyyne peak of TMC-1 and L1521E. 
We have detected radio spectral lines of various carbon-chain molecules 
such as CCS, C$_{3}$S, C$_{4}$H, HC$_{3}$N, and HC$_{5}$N. 
On the other hand, the NH$_{3}$ lines are weak and 
the N$_{2}$H$^{+}$ lines are not detected. 
According to our mapping observations of the HC$_{3}$N, CCS, and 
C$_{3}$S lines, the dense cores in L1495B and L1521B are 
compact with the radius of 0.063 and 0.044 pc, respectively, 
and have a simple elliptical structure. 
The distributions of CCS seem to be different 
from those of well-studied starless cores, L1498 and L1544, 
where the distribution of CCS shows a shell-like structure. 
Since the H$^{13}$CO$^{+}$, HN$^{13}$C, and C$^{34}$S lines are 
detected in L1495B and L1521B, the densities of these cores are high 
enough to excite the NH$_{3}$ and N$_{2}$H$^{+}$ lines. 
Therefore, the abundances of NH$_{3}$ and N$_{2}$H$^{+}$ relative to 
carbon-chain molecules are apparently deficient, as observed in L1521E. 
We found that longer carbon-chain molecules 
such as HC$_{5}$N and C$_{4}$H are more abundant in TMC-1 than L1495B 
and L1521B, while those of sulfur-bearing molecules such as 
C$^{34}$S, CCS, and C$_{3}$S are comparable. 
Both distributions and abundances of the observed molecules of 
L1495B and L1521B are quite similar to those of L1521E, strongly 
suggesting that L1495B and L1521B is in a very early stage of 
physical and chemical evolution. 
\end{abstract}


\section{Introduction}

It has been well established that dense cores in dark clouds are 
formation sites of low-mass stars. 
Because dense cores have low kinetic temperature (10 K) and 
high H$_{2}$ density ($>10^{3}$ cm$^{-3}$), 
they have been observed mainly with molecular lines in radio wavelength. 
Among a number of observational studies, the most systematic and 
extensive survey observations of NH$_{3}$ 
by Myers and his collaborators have greatly contributed to understanding 
of the physical property of dark cloud cores 
(e.g. Benson \& Myers 1989). 
They found that 68 \% of the dense cores observed by the NH$_{3}$ 
lines accompany the {\it{IRAS}} sources, 
which are newly born stars in the cores (Benson \& Myers 1989; 
Beichman et al. 1986). Therefore, it has been recognized that the NH$_{3}$ 
lines are useful to study the physical properties of star forming 
dense cores in dark clouds. 

On the other hand, 
Suzuki et al. (1992) pointed out that NH$_{3}$ is not always 
a good tracer of dense cores because of the chemical abundance 
variation from core to core. They carried out survey 
observations of CCS, HC$_{3}$N, HC$_{5}$N, and NH$_{3}$ 
toward 49 dense cores and found that the spectra of carbon-chain molecules 
tend to be intense in starless cores, 
while those of NH$_{3}$ tend to be intense in star forming cores. 
Especially, Suzuki et al. (1992) identified a few cores 
called "carbon-chain-producing regions", where the lines of 
carbon-chain molecules are intense while the NH$_{3}$ 
lines are hardly detected. 
They are L1495B, L1521B, L1521E, and 
the cyanopolyyne peak of TMC-1. 
Recently, Hirota, Ikeda, \& Yamamoto (2001, 2003) reported that 
deuterium fractionation ratios of DNC/HNC and 
DCO$^{+}$/HCO$^{+}$ are significantly lower in 
carbon-chain-producing regions than in the others. 
These systematic abundance variations 
would reflect the difference in chemical evolutionary stages of the cores; 
carbon-chain molecules and NH$_{3}$ are abundant in relatively 
early and late stages, respectively (Suzuki et al. 1992) 
and deuterium fractionation ratios increase as the core evolves 
(Hirota et al. 2001; Saito et al. 2000; Saito et al. 2002). 

Although detailed studies on carbon-chain-producing regions, 
which cannot be traced by the NH$_{3}$ lines, are 
essential for understanding of chemical and physical evolution 
of dense cores (Suzuki et al. 1992), 
no systematic study has been carried out for a long time, except for TMC-1 
(e.g. Olano, Walmsley, \& Wilson 1988; Hirahara et al. 1992; Pratap et al. 1997). 
In order to investigate the basic physical and chemical properties, 
we carried out detailed observations of a representative 
carbon-chain-producing region, L1521E, with various molecular lines 
(Hirota, Ito, \& Yamamoto 2002). 
The important results obtained there are as follows; 
(1) there exists a compact dense core traced by the 
H$^{13}$CO$^{+}$, HN$^{13}$C, CCS, and HC$_{3}$N lines, and 
their distributions have a single peak at the same position; 
(2) the distribution of CCS in L1521E is 
different from those in well-studied starless cores, L1498 and L1544, 
where the distribution of CCS shows a shell-like structure; 
(3) although the H$_{2}$ density is as high as 
(1.3-5.6)$\times$10$^{5}$ cm$^{-3}$ at the center of L1521E, 
the inversion lines of NH$_{3}$ are found to be 
very faint in L1521E, indicating the low NH$_{3}$ abundance; 
(4) abundances of carbon-chain molecules in L1521E are systematically 
higher than those in the other dark cloud cores, and especially 
the abundances of sulfur-bearing carbon-chain 
molecules C$_{n}$S are comparable to 
those in cyanopolyyne peak of TMC-1. 
According to these results along with a fact that there exist 
neither {\it{IRAS}} point sources, evidence of molecular outflow, 
nor signature of infall motion, 
we suggested that L1521E would be in 
a very early stage of physical and chemical evolution 
(Hirota et al. 2002).  
Very recently, detailed chemical model calculations 
(Aikawa, Ohashi, \& Herbst 2003) and molecular line and millimeter continuum observations 
(Tafalla \& Santiago 2004) also confirmed our results. 

The carbon-chain-producing regions would be chemically and 
dynamically less evolved than other dark cloud cores, 
and are rare objects. In fact, only 4 such sources have been recognized 
(Suzuki et al. 1992). 
Therefore, it is important to carry out detailed studies on 
the other carbon-chain-producing regions in order to understand 
general properties of dense cores in a very early stage of 
chemical and dynamical evolution. 
Although L1521B has been observed and detected in previous 
molecular line surveys (NH$_{3}$ for Benson \& Myers 1989; 
HC$_{5}$N for Benson \& Myers 1983; CCS for Ohashi 2000; 
H$^{13}$CO$^{+}$ Onishi et al. 2002), L1495B was not detected 
in such survey observations except for Suzuki et al. (1992) and 
Hirota et al. (1998; 2001). 
In this paper, we report observations of L1495B and L1521B with 
various molecular lines. 

\section{Observations}

The observed lines are summarized in Table \ref{tab-observe}. 
We took the reference position of L1495B to be 
$\alpha_{1950}=04^{h}12^{m}30^{s}.0$, 
$\delta_{1950}=28^{\circ}$39\arcmin39\arcsec. 
Note that this reference position is taken to be 1\arcmin \ north 
of that in the previous papers (e.g. Myers, Linke, \& Benson 1983; 
Suzuki et al. 1992). 
We took the reference position of L1521B to be 
$\alpha_{1950}=04^{h}21^{m}08^{s}.5$, 
$\delta_{1950}=26^{\circ}$30\arcmin00\arcsec, 
which is the same as most of previous papers 
(e.g. Myers et al. 1983; Suzuki et al. 1992). 
 
\placetable{tab-observe}

The 42-45 GHz and 93 GHz lines were observed with the 45 m radio 
telescope at Nobeyama Radio Observatory (NRO)\footnotemark 
\footnotetext{Nobeyama Radio Observatory is a branch of 
the National Astronomical Observatory of Japan, an interuniversity 
research institute operated by the Ministry of 
Education, Science, Sports and Culture of Japan} 
in several observing sessions from 1990 to 1999. 
All of them were observed with SIS mixer 
receivers whose system temperatures were 200-500 K. 
The main-beam efficiencies ($\eta_{mb}$) were 0.7 and 0.5 for the 
42-45 GHz and 93 GHz regions, respectively, and the beam sizes 
were 37\arcsec \ and 17\arcsec \ for the 42-45 GHz and 
93 GHz regions, respectively. 
Acousto-optical radio spectrometers 
with the frequency resolution of 37 kHz were 
used for the backend. 
Pointing was checked by observing a nearby SiO maser source, 
NML-Tau, every 1-2 hours, and the pointing accuracy was 
estimated to be better than 5\arcsec. 
All the observations were performed with the 
position-switching mode, in which a typical off position was 10\arcmin \ 
away from the source position. 
In the mapping observations of HC$_{3}$N and CCS, the spectra 
were usually observed with a grid spacing of 60\arcsec. The spectra were 
taken with a finer grid spacing of 30\arcsec \ toward 
the central part of the cores. 

The 19-23 GHz lines 
were observed with the Effelsberg 100 m radio telescope of 
Max-Planck-Institut f\"ur Radioastronomie in 2000 May. 
We used cooled HEMT receivers, whose system temperatures 
were about 30 K. 
The digital autocorrelators with the frequency resolution of 10 kHz 
were used for the backend. 
For pointing and intensity calibrations, we observed 3C123 every 1-2 hours. 
We adopted a flux density of 3C123 to be 3.12 Jy 
at 23780 MHz (Ott et al. 1994). 
Observations were carried out with the frequency switching mode, 
in which the offset frequency was set to be 0.2 MHz. 
In the mapping observations of C$_{3}$S, the grid spacing is taken to be 40\arcsec. 

\section{Results and Discussions}

\subsection{Mapping Observations of L1495B and L1521B} 

\begin{figure}
\epsscale{1}
\plotone{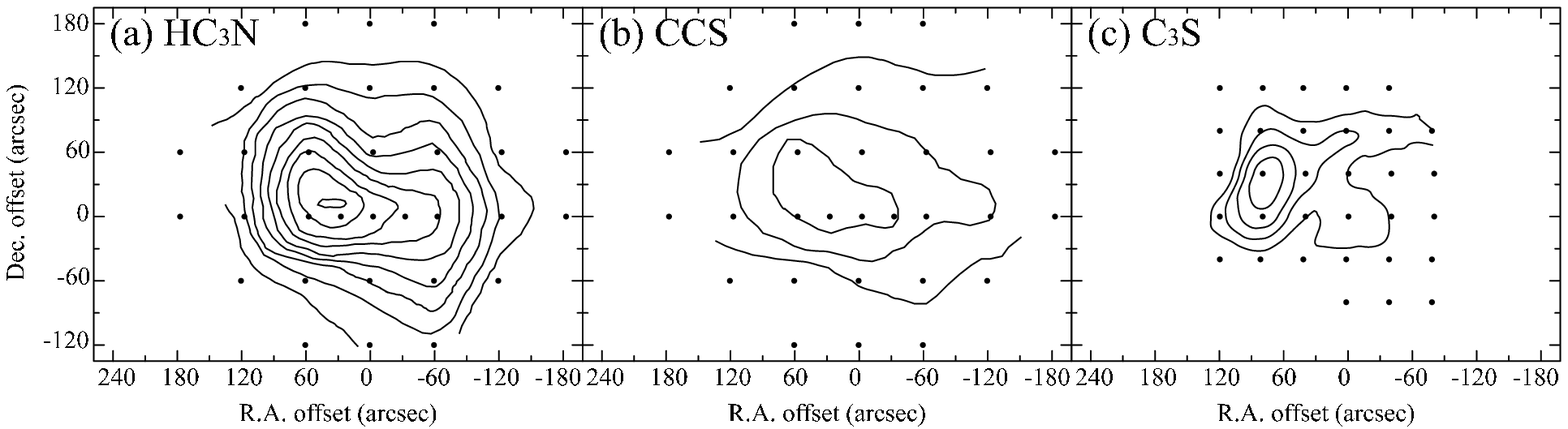}
\caption{Integrated intensity maps of the observed molecules 
toward L1495B. The reference position is 
$\alpha_{1950}=04^{h}12^{m}30^{s}.0$, 
$\delta_{1950}=28^{\circ}$39\arcmin39\arcsec. 
(a) HC$_{3}$N ($J=5-4$). 
The velocity range of integration is from 7.1 to 8.1 km s$^{-1}$. 
The interval of the contours is 0.27 K km s$^{-1}$ and 
the lowest one is 0.27 K km s$^{-1}$. 
(b) CCS ($J_{N}=4_{3}-3_{2}$). 
The velocity range of integration is from 7.3 to 7.8 km s$^{-1}$. 
The interval of the contours is 0.21 K km s$^{-1}$ and 
the lowest one is 0.21 K km s$^{-1}$. 
(c) C$_{3}$S ($J=4-3$). 
The velocity range of integration is from 7.0 to 8.0 km s$^{-1}$. 
The interval of the contours is 0.07 K km s$^{-1}$ and 
the lowest one is 0.11 K km s$^{-1}$. 
\label{fig-l1495bmaps}  }
\end{figure}

\begin{figure}
\epsscale{1}
\plotone{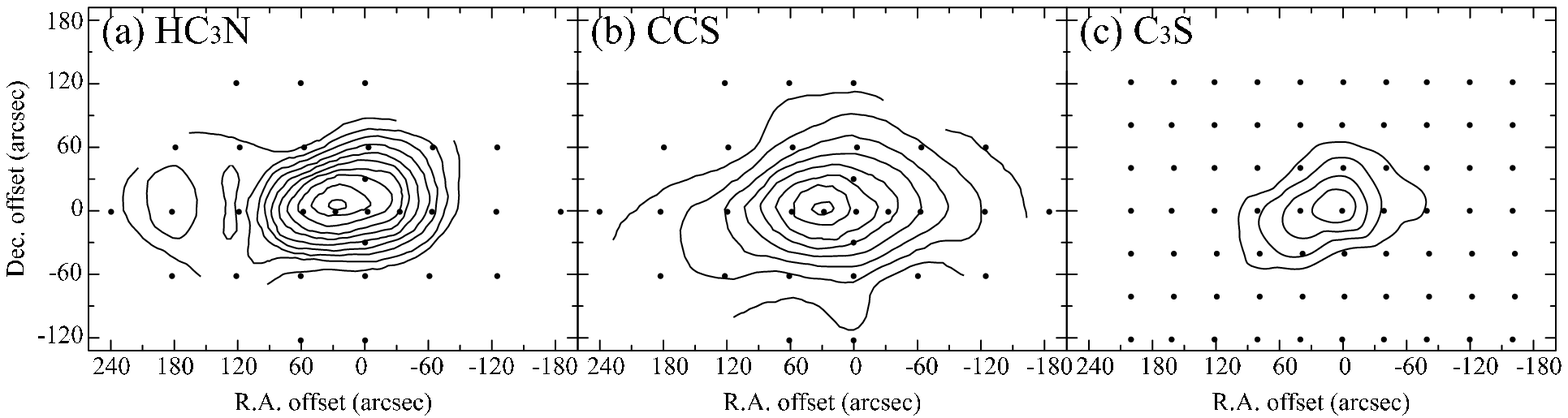}
\caption{Integrated intensity maps of the observed molecules 
toward L1521B. The reference position is 
$\alpha_{1950}=04^{h}21^{m}08^{s}.5$, 
$\delta_{1950}=26^{\circ}$30\arcmin00\arcsec. 
(a) HC$_{3}$N ($J=5-4$). 
The velocity range of integration is from 5.5 to 7.0 km s$^{-1}$. 
The interval of the contours is 0.27 K km s$^{-1}$ and 
the lowest one is 0.27 K km s$^{-1}$. 
(b) CCS ($J_{N}=4_{3}-3_{2}$). 
The velocity range of integration is from 6.0 to 6.8 km s$^{-1}$. 
The interval of the contours is 0.21 K km s$^{-1}$ and 
the lowest one is 0.21 K km s$^{-1}$. 
(c) C$_{3}$S ($J=4-3$). 
The velocity range of integration is from 6.0 to 7.0 km s$^{-1}$. 
The interval of the contours is 0.15 K km s$^{-1}$ and 
the lowest one is 0.23 K km s$^{-1}$. 
\label{fig-l1521bmaps}   }
\end{figure}

Figures \ref{fig-l1495bmaps} and \ref{fig-l1521bmaps} show 
the integrated intensity maps of HC$_{3}$N, CCS, and 
C$_{3}$S toward L1495B and L1521B, respectively. 
It is clearly found that there exists a compact dense core traced by these 
lines, which has a single peak and a simple elliptical shape. 
For L1495B, the peak position of HC$_{3}$N and CCS is 
located at ($\Delta \alpha, \Delta \delta$)=(60\arcsec, 0\arcsec), whereas 
the C$_{3}$S peak is at (80\arcsec, 40\arcsec). 
This difference may originate from the difference 
in grid spacing; the grid spacing are 30\arcsec \ for HC$_{3}$N and CCS, 
and 40\arcsec \ for C$_{3}$S. 
For L1521B, the peak position of HC$_{3}$N, CCS, and C$_{3}$S 
is located at (30\arcsec, 0\arcsec). This position lies close to 
the northwest peak position of the H$^{13}$CO$^{+}$ 
core MC 22 observed by Onishi et al. (2002). 
The simple elliptical distribution of CCS in L1521B is also reported by 
Ohashi (2000), in which they observed the $3_{2}-2_{1}$ line of CCS in the 
33 GHz band with the BIMA array. 

The single peak distribution of CCS observed in L1495B and L1521B, 
as well as L1521E (Hirota et al. 2002), is not a usual feature in 
starless cores. 
For L1498 (Kuiper, Langer, \& Velusamy 1996), L1544 
(Ohashi et al. 1999; Benson \& Myers 1989), and several other cores 
(Ohashi 2000; Lai \& Crutcher 2000), the distribution of 
CCS shows a shell-like structure, while that of NH$_{3}$ is concentrated 
at the central position (e.g. Tafalla et al. 2002). 
Such distributions can be interpreted qualitatively 
in terms of the chemical evolution; CCS tends to be deficient at the center of 
chemically evolved cores, whereas NH$_{3}$ becomes abundant there 
(Suzuki et al. 1992; Bergin \& Langer 1997; Aikawa et al. 2003). 
In contrast to L1498 and L1544, CCS is not deficient at the center of 
the core in L1495B, L1521B, and L1521E. 

Assuming that the distance to the Taurus Molecular Cloud 
is 140 pc (Elias 1978), the core radius is derived to be 0.063 pc and 0.044 pc 
for L1495B and L1521B, respectively, 
from two dimensional Gaussian fitting of the integrated intensity map of CCS. 
The core radius of L1521B is comparable to that derived from 
the H$^{13}$CO$^{+}$ map, 0.054 pc (Onishi et al. 2002). 
These size are comparable to those of typical starless NH$_{3}$ cores 
(0.12$\pm$0.06 pc in diameter; Benson \& Myers 1989), 
starless N$_{2}$H$^{+}$ cores (0.05$\pm$0.02 in radius; Caselli et al. 2002), 
and H$^{13}$CO$^{+}$ cores in the Taurus Molecular Cloud 
(0.048$\pm$0.023 pc in radius; Onishi et al. 2002), 
while they are 1.5-2 times larger than that of another 
carbon-chain-producing region L1521E, 0.031 pc (Hirota et al. 2002). 

The total mass of a core, $M_{tot}$, is derived by the following equation; 
\begin{equation}
M_{tot} = 0.24 r^{3} n(\mbox{H}_2) (M_{\odot})
\end{equation}
where $r$ and $n($H$_{2})$ are the core radius in pc and the peak 
H$_{2}$ density in cm$^{-3}$, respectively. 
Here we assume that the cores are homogeneous spheres with constant 
density. Because the density profile is not uniform in starless cores, 
(e.g. Ward-Thompson et al. 1999; 
Evans et al. 2001; Tafalla et al. 2002, 2004), 
the mass presented here would be overestimated. 
The peak H$_{2}$ density in L1521B is derived to be 
1.9$\times$10$^{5}$ cm$^{-3}$ and 
(6-7)$\times$10$^{4}$ cm$^{-3}$ from multi-transition observations of 
C$^{34}$S (Hirota et al. 1998) and CCS (Suzuki et al. 1992), respectively. 
If we assume the radius of 0.044 pc and the H$_{2}$ density of 
7$\times$10$^{4}$ cm$^{-3}$, the total mass of 
L1521B is calculated to be 1.4$M_{\odot}$. 
If the radius of 0.054 pc and the H$_{2}$ density of 
1.9$\times$10$^{5}$ cm$^{-3}$ are assumed, the total mass is estimated 
to be 7.2$M_{\odot}$. 
Thus the mass of L1521B ranges from 1.4$M_{\odot}$ to 7.2$M_{\odot}$, 
being comparable to that reported by 
Onishi et al. (2002), 5.9$M_{\odot}$. 

The total mass of L1495B is derived to be 4.2$M_{\odot}$, 
assuming the radius of 0.063 pc and the H$_{2}$ density of 
7$\times$10$^{4}$ cm$^{-3}$. 
Although the uncertainty in the total mass of L1495B and L1521B 
is estimated to be a factor of 3, they are comparable to that of 
L1521E, 2.4$M_{\odot}$ (Hirota et al. 2002), and correspond to 
the lower end of the mass of starless NH$_{3}$ cores, 
16$\pm$30 $M_{\odot}$ (Benson \& Myers 1989). 
The average value of the mass of starless N$_{2}$H$^{+}$ cores found 
by Caselli et al. (2002) is 3$M_{\odot}$, similar to those found for 
L1495B and L1521B.

The virial mass of a core, $M_{vir}$, is derived by the following equation; 
\begin{eqnarray}
M_{vir} & = & 210 r \Delta v_{m}^{2} (M_{\odot}) \\ 
\Delta v_{m}^{2} & = & \Delta v^{2}  
  - \Delta v_{res}^{2} 
  + 0.461\left(\frac{1}{m_{mean}}-\frac{1}{m_{X}} \right)
\end{eqnarray}
where $\Delta v$ is the observed linewidth in km s$^{-1}$, 
$\Delta v_{res}$ is the velocity resolution of the spectrometer in km s$^{-1}$, 
$\Delta v_{m}$ is the total linewidth of the molecule of mean mass 
including both nonthermal and 
thermal motion in the kinetic temperature of 10 K, 
$m_{mean}$ is the mean mass of molecules in atomic mass unit, 
2.33 amu, and  $m_{X}$ is the mass of the observed molecules in 
atomic mass unit. 
The total linewidths derived from the CCS lines are 
0.48 km s$^{-1}$ and 0.54 km s$^{-1}$ for L1495B and L1521B, 
respectively, and hence, the virial masses are estimated to be 
3.0$M_{\odot}$ and 2.7$M_{\odot}$, respectively. 
If we adopt the linewidths derived from the H$^{13}$CO$^{+}$ lines, 
the total linewidths are 0.62 km s$^{-1}$ and 0.69 km s$^{-1}$ 
for L1495B and L1521B, respectively, and the virial masses are 
estimated to be 5.0$M_{\odot}$ and 4.4$M_{\odot}$, respectively. 

Although both the total mass and the virial mass contain large 
uncertainty of typically a factor of 3, L1495B and L1521B are 
not likely to be unstable against collapse, but rather in virial equilibrium. 
It is consistent with the fact that evidences of infalling motions 
(Lee, Myers, \& Tafalla 1999) 
have never been reported for L1521B, suggesting at least L1521B 
is not in a dynamically collapsing phase. 
The {\it{IRAS}} point sources are not associated with L1495B 
and L1521B, and molecular outflows 
have never been reported for L1495B and L1521B. 
Hence, star formation activities have not yet started in L1495B and L1521B. 
These facts suggest that the L1495B and L1521B cores are in the 
early stage of dynamical evolution. 
However, the detailed physical properties of L1495B and L1521B 
should be further investigated with other well-studied high density 
tracers such as the C$^{18}$O and H$^{13}$CO$^{+}$ lines 
(Myers et al. 1983; Onishi et al. 2002) 
and submillimeter continuum emission in order to compare the dynamical 
evolutionary stages of L1495B and L1521B with those of other dark cloud cores
(e.g. Tafalla \& Santiago 2004; Tafalla et al. 2002, 2004). 
In the present observations, 
we could not find significant velocity gradient across the cores 
nor significant variation of linewidths within the cores 
which are observed in several more evolved starless cores 
(e.g. Caselli et al. 2002; Tafalla et al. 2004), partly because of 
insufficient spectral resolution of our observations. 
On the other hand, linewidths of the H$^{13}$CO$^{+}$ lines in 
L1495B and L1521B are comparable to those found in 
more evolved starless cores such as L1521F (Onishi et al. 2002). 
Further observations with high spectral resolution would be necessary for 
detailed understandings of velocity structures of L1495B and L1521B. 

\placefigure{fig-l1495bmaps}
\placefigure{fig-l1521bmaps}

\subsection{Molecular Abundances of L1495B and L1521B}

\begin{figure}
\epsscale{0.6}
\plotone{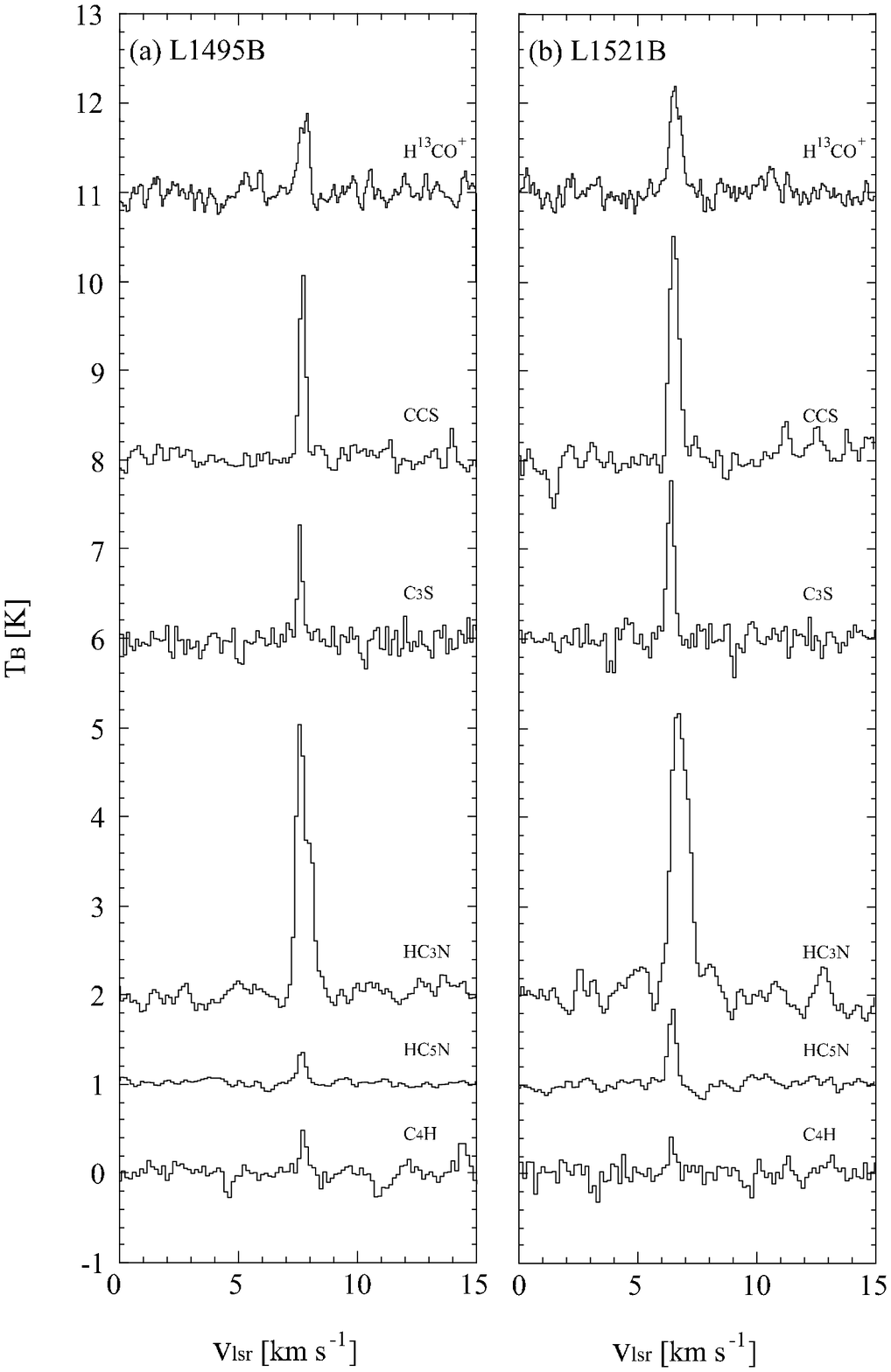}
\caption{Spectra of H$^{13}$CO$^{+}$, 
CCS, C$_{3}$S, HC$_{3}$N, HC$_{5}$N, and C$_{4}$H. 
These spectra are observed toward the reference position of 
(a) L1495B and (b) L1521B except for C$_{3}$S and C$_{4}$H toward L1495B, which are 
observed at the (80\arcsec, 40\arcsec) position corresponding to 
the C$_{3}$S peak. 
The apparent features below the baseline of the C$_{3}$S and 
C$_{4}$H spectra are artifacts of the frequency switching technique. 
The hyperfine components of C$_{4}$H can be seen at 
$v_{lsr}$ of 7.7 km s$^{-1}$ ($F$=3-2) and 14.4 km s$^{-1}$ 
($F$=2-1) for L1495B, while only one component at 
$v_{lsr}$ of 6.4 km s$^{-1}$ ($F$=3-2) is detected for L1521B. 
\label{fig-spccs} }
\end{figure}

The top frame of Figure \ref{fig-spccs} shows the spectra of 
H$^{13}$CO$^{+}$($J$=1-0) 
observed toward the reference positions of L1495B and L1521B. 
The peak brightness temperature of the H$^{13}$CO$^{+}$ 
line is similar to those observed toward other H$^{13}$CO$^{+}$ cores 
(Onishi et al. 2002). Although we did not carry out statistical equilibrium 
calculations to evaluate the H$_{2}$ density because of the lack of 
multi-transition observations, detection of the H$^{13}$CO$^{+}$ ($J$=1-0) 
line is an evidence of dense core with the H$_{2}$ density 
of an order of 10$^{5}$ cm$^{-3}$. 
In fact, the H$_{2}$ density derived from the C$^{34}$S data in 
L1521B is 1.9$\times$10$^{5}$ cm$^{-3}$ (Hirota et al. 1998). 

\placefigure{fig-spccs}

\begin{figure}
\epsscale{0.6}
\plotone{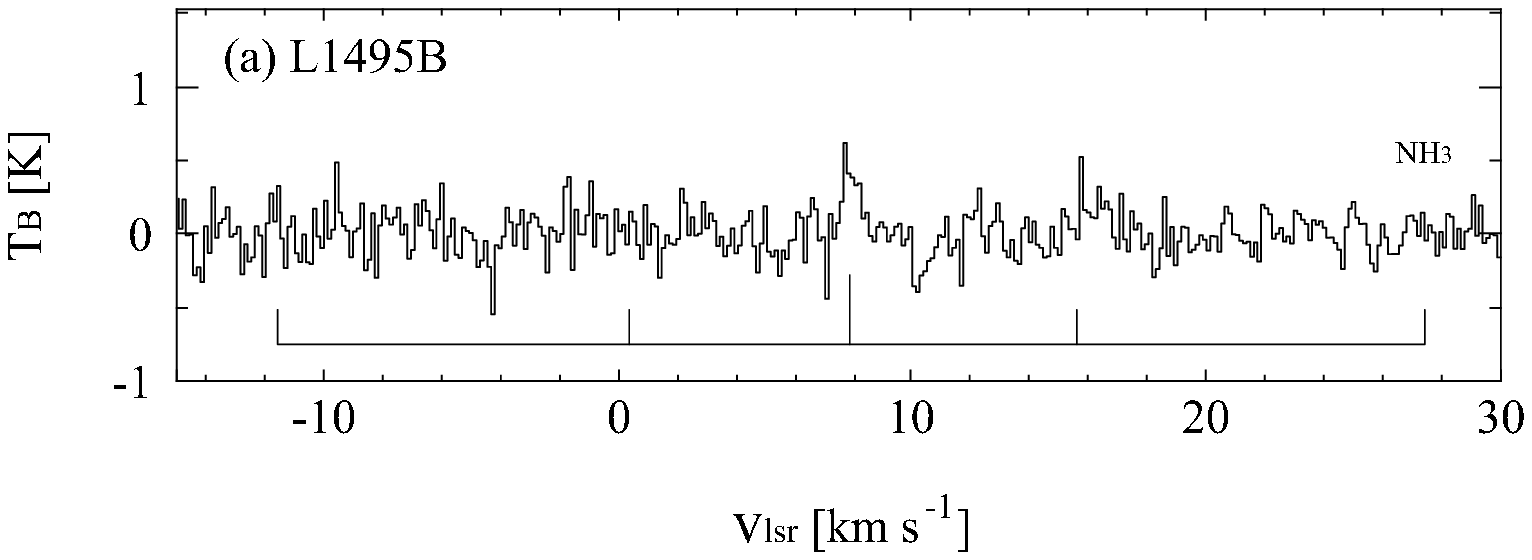}
\plotone{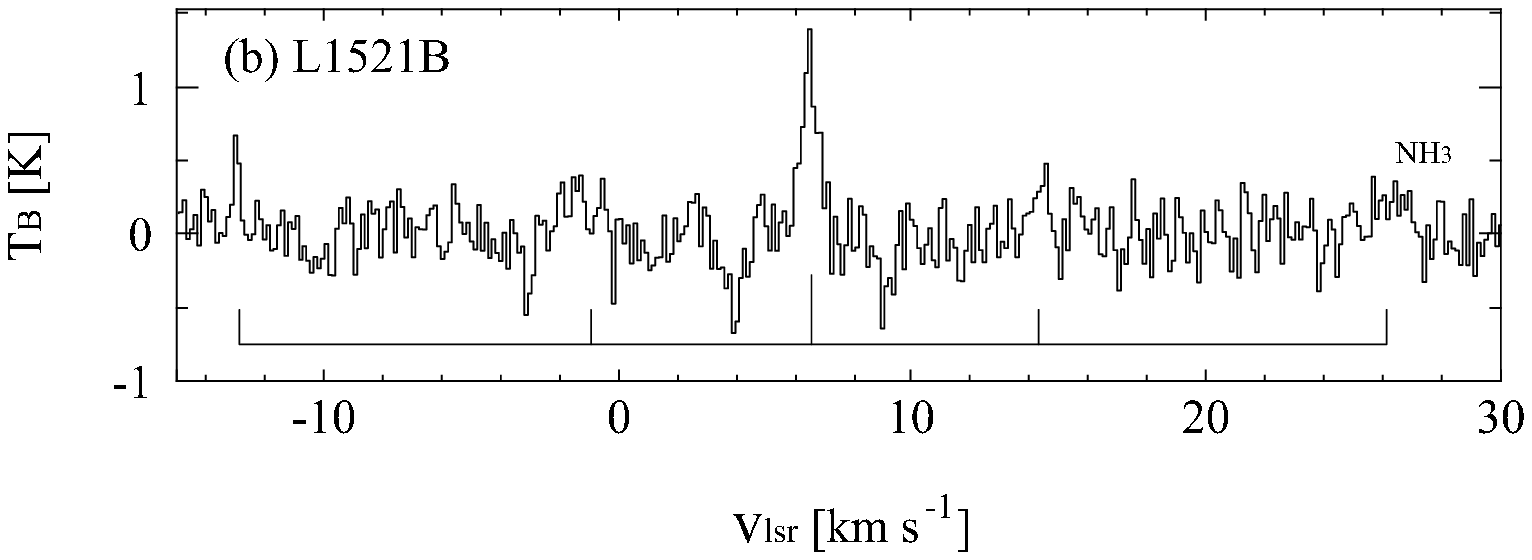}
\caption{Spectra of NH$_{3}$($J,K$=1,1) 
at the reference position of (a)L1495B and (b)L1521B. 
The apparent features below the baseline of the NH$_{3}$ 
spectra are artifacts of the frequency switching technique. 
The expected positions of hyperfine components of the NH$_{3}$ 
line are indicated in the bottom of each figure. 
\label{fig-spnh3} }
\end{figure}

Although the H$_{2}$ density is high enough to excite the inversion 
line of NH$_{3}$($J, K$=1,1), it is found to be faint toward 
the reference position of L1495B and L1521B, as shown in Figure \ref{fig-spnh3}. 
According to the survey of the NH$_{3}$ lines by Benson \& Myers (1989) and 
Suzuki et al. (1992), the NH$_{3}$ line was not detected toward L1495B. 
For L1521B, the NH$_{3}$ line was detected with the brightness temperature of 
0.48 K (Benson \& Myers 1989) and 0.36 K (Suzuki et al. 1992). 
Owing to our higher sensitivity observations with higher spatial resolution, 
the observed brightness temperature of the NH$_{3}$ lines, 
0.49 K and 0.94 K for L1495B and L1521B, respectively, are higher by a factor 
of 2 or more than those reported by Benson \& Myers (1989) and 
Suzuki et al. (1992). 
Even with this results, the brightness temperature of the NH$_{3}$ lines toward 
L1495B and L1521B fall within 
the range of the criterion of "moderate" detection by the survey of 
Benson \& Myers (1989), 1.4 K $> T_{B}\equiv T_{A}^{*}/\eta_{b} >$ 3$\sigma$. 
This result indicates that abundances of NH$_{3}$ are quite low at least in 
the peak position of dense cores in L1495B and L1521B traced by 
the HC$_{3}$N, CCS, and C$_{3}$S lines. 
Such characteristics are similar to those in another 
carbon-chain-producing region L1521E (Hirota et al. 2002). 

\placefigure{fig-spnh3}

Figure \ref{fig-spccs} shows the spectra of carbon-chain molecules 
observed toward L1495B and L1521B. The observed position is 
the reference position except for C$_{3}$S and C$_{4}$H toward L1495B, 
which are observed at the C$_{3}$S peak. 
In addition to the intense spectra of CCS and HC$_{3}$N, 
the longer carbon-chain molecules such as C$_{3}$S, C$_{4}$H, and 
HC$_{5}$N were also detected toward both L1495B and L1521B, 
indicating that L1495B and L1521B are rich in carbon-chain-molecules. 
In order to evaluate abundances of observed molecules, 
the line parameters for all the observed lines were determined by the 
Gaussian fit, as summarized in Table \ref{tab-observe}. 
We ignored the unresolved hyperfine 
structures of the NH$_{3}$ and HC$_{3}$N lines, and hence, 
their linewidths are broader than the others. 

We calculated the column densities of observed molecules 
by the consistent way employed in 
Hirota et al. (2002) in order to compare the present results 
with those for L1521E and TMC-1. Details of our method are 
described in Hirota et al. (2002). 
We assumed that the excitation temperatures of 
6.5 K for NH$_{3}$, HC$_{3}$N, and HC$_{5}$N, 
6.0 K for C$_{4}$H, and 5.5 K for C$_{3}$S. 
For NH$_{3}$, all the ortho and para levels 
were assumed to be thermalized at the kinetic temperature of 10 K. 
The upper limit of the column density of N$_{2}$H$^{+}$ is calculated 
by assuming the excitation temperature of 5.0 K 
(Benson, Caselli, \& Myers 1998) and the FWHM linewidth of 0.5 km s$^{-1}$. 
The derived column densities are 
summarized in Table \ref{tab-column} along with those of several 
molecules reported previously. 

For L1521B, the column densities of C$_{3}$S, HC$_{5}$N, and NH$_{3}$  
are reported by Suzuki et al. (1992) at the same position. 
The column densities of C$_{3}$S and NH$_{3}$ obtained by 
our observations are factor of 2 larger than those of Suzuki et al. (1992), 
possibly because we observed the C$_{3}$S and NH$_{3}$ lines with 
a finer beam size and our results are less affected by the beam dilution effect. 
For the HC$_{5}$N line, we observed 
the $J$=16-15 line while Suzuki et al. (1992) observed the 
$J$=17-16 line with the same telescope and a similar beam size. 
The column density of HC$_{5}$N obtained in the present study is 
1.2$\times$10$^{13}$ cm$^{-2}$, which well agrees to 
that reported by Suzuki et al. (1992), 1.17$\times$10$^{13}$ cm$^{-2}$. 

For L1495B, two hyperfine components of the $N$=2-1 transition of 
C$_{4}$H ($F$=3-2 and 2-1) are detected. The intensity ratio of 
$T_{B}(F=3-2)$/$T_{B}(F=2-1)$ is 1.32$\pm$0.54, agrees to the 
intensity ratio in the optically thin case 
(Gu\'elin, Friberg, \& Mezaoui 1982).  
Therefore, we calculated column density of C$_{4}$H using 
one of the $F$=3-2 and $F$=2-1 data. 
Assuming the same excitation temperature of 6.0 K 
as adopted in the analysis of L1521E (Hirota et al. 2002) and L1521B, 
the total optical depth of the C$_{4}$H lines are derived to be 
0.47 and 0.51 from the $F$=3-2 and $F$=2-1 data, respectively, 
and the column density of C$_{4}$H are derived to be 
1.5$\times$10$^{14}$ cm$^{-2}$ and 2.1$\times$10$^{14}$ cm$^{-2}$ 
from the $F$=3-2 and $F$=2-1 data, respectively. 
Therefore, we adopted the average value derived 
from the $F$=3-2 and $F$=2-1 data, 1.82$\times$10$^{14}$ cm$^{-2}$ 
as the final value. 

\placetable{tab-colmn}

The column densities of the carbon-chain molecules in L1495B and L1521B 
are significantly higher than those in typical dark cloud cores  
(Benson \& Myers 1983; Fuente et al. 1990; Suzuki et al. 1992), 
whereas the column densities of NH$_{3}$ and N$_{2}$H$^{+}$ 
in L1495B and L1521B are lower by a factor of 5-10 
(Benson \& Myers 1989; Suzuki et al. 1992; Benson et al. 1998; 
Caselli et al. 2002). 
The NH$_{3}$/CCS ratios in carbon-chain-producing regions, 
ranging from 2.6 (L1521E) to 3.8 (L1495B), are lower 
by a factor of 10 or more than those in other dark cloud cores 
(e.g. Figure 7 of Ohishi \& Kaifu 1998, Figure 10 of Hirota et al. 2001). 
The abundances of the carbon-chain-molecules relative to 
NH$_{3}$ and N$_{2}$H$^{+}$ 
are systematically higher in L1495B and L1521B than those in 
typical dark cloud cores. 

\begin{figure}
\epsscale{0.5}
\plotone{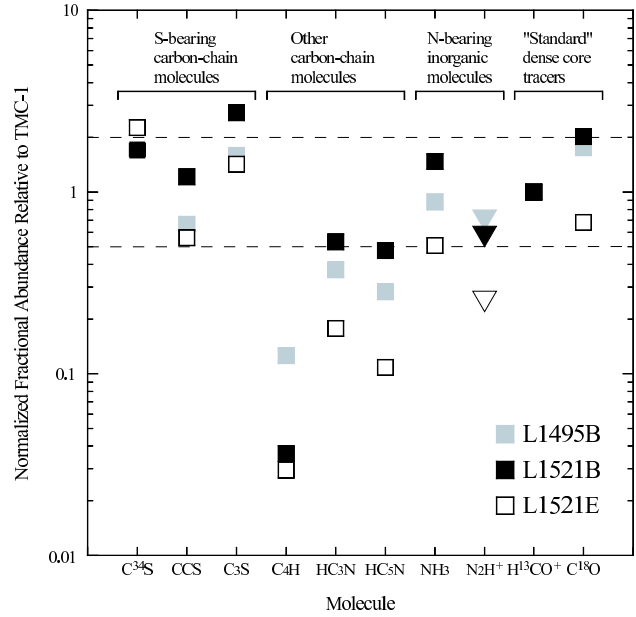}
\caption{"Normalized" fractional abundances of selected molecules 
relative to H$^{13}$CO$^{+}$. 
The values are normalized to those of TMC-1. 
The value of C$^{34}$S in L1495B is not plotted. 
Triangles for N$_{2}$H$^{+}$ indicate upper limits. 
Horizontal dashed lines indicate the difference in the normalized 
fractional abundance relative to TMC-1 is a factor of 2. 
Because the fractional abundance are calculated relative to 
H$^{13}$CO$^{+}$, the values for H$^{13}$CO$^{+}$ are 1 for 
all the 3 sources. 
\label{fig-abundance} }
\end{figure}

In order to illustrate how the molecular abundances 
are different among carbon-chain-producing regions, 
we show the fractional abundances of 
selected molecules relative to the TMC-1 abundances 
("normalized" fractional abundances) in Figure \ref{fig-abundance}. 
Because the volume of gas traced by CO and its isotopes are not 
always the same as those of other molecules due to their low 
critical density, we compared the molecular abundances relative to 
H$^{13}$CO$^{+}$ in this paper, following the discussions of 
TMC-1 by Pratap et al. (1997) and of L1521E by Hirota et al. (2002). 
Aikawa et al. (2003) recently reported chemical model calculations 
describing that column density of 
HCO$^{+}$ does not show temporal variation unless the H$_{2}$ density 
is less than 10$^{5}$ cm$^{-3}$, and hence, it is useful as 
a "standard" dense core tracer. 
For example, the normalized fractional abundance of CCS in L1495B 
relative to that in TMC-1, $X$(CCS), is defined as follows; 
\begin{equation}
X\mbox{(CCS)}=\frac{[\mbox{CCS(L1495B)}]/[\mbox{H}^{13}\mbox{CO}^{+}\mbox{(L1495B)}]}
{[\mbox{CCS(TMC-1)}]/[\mbox{H}^{13}\mbox{CO}^{+}\mbox{(TMC-1)}]}.
\end{equation}
According to the systematic abundance variation of molecular species in 
L1521E (Hirota et al. 2002), we classified 4 groups of molecular species 
in Figure \ref{fig-abundance}; 
(1) sulfur-bearing carbon-chain molecules, C$^{34}$S, CCS, and C$_{3}$S, 
which are systematically higher in L1521E than TMC-1, 
(2) other carbon-chain molecules, C$_{4}$H, HC$_{3}$N, and HC$_{5}$N, 
which are systematically lower in L1521E than TMC-1, 
(3) nitrogen-bearing inorganic molecules, NH$_{3}$ and N$_{2}$H$^{+}$, 
and (4) "standard" dense core tracers, H$^{13}$CO$^{+}$ and C$^{18}$O. 

As observed in L1521E, fractional abundances of 
sulfur-bearing carbon-chain molecules are comparable to or higher in 
L1495B, L1521B, and L1521E than in TMC-1, while those of the other carbon-chain 
molecules are systematically lower by a factor of 2-30 
in all the 3 carbon-chain-producing regions than in TMC-1. 
On the other hand, the abundances of NH$_{3}$ and C$^{18}$O 
in all the 3 carbon-chain-producing regions are almost the same as TMC-1. 
Although the abundances of N$_{2}$H$^{+}$ are derived only with 3$\sigma$ 
upper limit, they seem to be systematically lower 
in all 3 carbon-chain-producing regions than in TMC-1. 
Therefore, chemical compositions of L1495B and L1521B are quite similar to 
that of L1521E (Hirota et al. 2002). 

According to a simple gas-phase pseudo-time dependent chemical model 
(e.g. Suzuki et al. 1992), 
CCS and carbon-chain molecules are abundant only in the early evolutionary 
stage of dark cloud cores. This is because CCS and carbon-chain molecules 
are efficiently produced when the C$^{+}$ ions and the C atoms are abundant 
while they are deficient when the C atoms are locked into CO. 
In addition, the CCS molecules freeze-out onto grains as the 
core density increases because sulfur-bearing species are tightly bound 
in grain mantles (Bergin \& Langer 1997). 
Therefore, we can conclude that L1495B, L1521B, and L1521E are in almost 
the same chemical evolutionary stage, and would be in the earliest phase of 
chemical evolution. 
On the other hand, Ruffle et al. (1997) presented completely different 
chemical model calculations suggesting that abundances of cyanopolyynes 
may increase at the late time where molecules freeze-out onto dust grains. 
However, it is well known both theoretically and observationally that the 
freeze-out of CO boosts the production rate of NH$_{3}$ and N$_{2}$H$^{+}$ 
(Bergin \& Langer 1997; Aikawa et al. 2003; Tafalla et al. 2002, 2004). 
This is not consistent with our observational result that the abundances of 
NH$_{3}$ and N$_{2}$H$^{+}$ in these three cores are lower than 
in other dark cloud cores. 
In addition, the freeze-out of CO makes the lifetime of H$_{2}$D$^{+}$ 
longer, and hence, deuterium fractionation ratios such as 
DCO$^{+}$/HCO$^{+}$ ratios are expected to be enhanced through the deuteron 
transfer reaction from H$_{2}$D$^{+}$ followed by the electron recombination
reaction (Saito et al. 2002; Aikawa et al. 2003). 
Again, this is not consistent with the low DCO$^{+}$/H$^{13}$CO$^{+}$ ratios 
observed in L1495B, L1521B, and L1521E (Hirota et al. 2001). 
From these considerations, it is likely that freeze-out of CO 
and other molecules onto dust grain are not efficient in these two cores 
as observed by Tafalla \& Santiago (2004). 

It will be interesting to compare typical time scales for dynamical 
and chemical evolution. 
The timescale for gas-phase chemical reactions is roughly estimated 
to be 10$^{5}$ yr for an H$_{2}$ density of 10$^{5}$ cm$^{-3}$ and 
a cosmic ray ionization rate of 10$^{-17}$ s$^{-1}$. 
This timescale is comparable to the free-fall  
timescale (10$^{5}$ yr) and the timescale of freeze-out of CO molecules 
onto dust grain (10$^{5}$ yr). 
The fact that the freeze-out of molecules is not significant in 
L1495B and L1521B suggests that these two cores are dynamically young. 
Therefore, we can conclude that L1495B and L1521B are really in the earliest 
phase of not only chemical evolution 
but also physical evolution, as in the case of L1521E 
(Hirota et al. 2002; Tafalla \& Santiago 2004). 
Although it is difficult to infer the ages of L1495B and L1521B quantitatively, 
they are likely to be less than 10$^{5}$ yr. Detailed comparison with chemical 
models including dynamical evolution (e.g. Aikawa et al. 2003) would be interesting.  

Recently, Lee et al. (2003) found chemically young but physically 
evolved core (L1689B) and chemically evolved but physically young core (L1512), 
pointing out that environmental conditions have to be taken into account 
when modeling the chemistry and physics of molecular cloud cores.  
In addition, initial condition of dynamical collapse and timescale of 
dynamical collapse should be considered in the model of dark cloud cores 
because these differences would affect the chemical abundance of dark cloud 
cores (e.g. Aikawa et al. 2003). 
In order to better outline the physical and chemical properties 
of L1495B and L1521B, submillimeter continuum observations to map the 
dust distributions and further detailed molecular line observations 
(e.g. Ward-Thompson et al. 1999; Tafalla et al. 2002, 2004; 
Tafalla \& Santiago 2004) 
are needed, which would contribute to the complete understanding of 
chemical and physical evolution of dark cloud cores. 

\placefigure{fig-abundance}

\section{Summary}

We carried out mapping and survey observations 
with several molecular lines toward two known 
carbon-chain-producing regions in the Taurus Molecular Cloud, 
L1495B and L1521B. 
Intense spectra of carbon-chain molecules 
such as CCS, C$_{3}$S, C$_{4}$H, HC$_{3}$N, and HC$_{5}$N were detected, 
while the NH$_{3}$ lines were weak and the N$_{2}$H$^{+}$ lines 
were not detected. 
Distributions of HC$_{3}$N, CCS, and C$_{3}$S lines in 
L1495B and L1521B show compact elliptical structure with 
the radius of 0.063 and 0.044 pc, respectively. 
The maps of the CCS lines has only single peak position and 
they seem to be different from those of well-studied starless 
cores, L1498 and L1544, 
where the distribution of CCS shows a shell-like structure. 
Since the H$^{13}$CO$^{+}$ lines are 
detected in L1495B and L1521B, the densities of these cores are 
higher than 10$^{5}$ cm$^{-3}$, which is high 
enough to excite the NH$_{3}$ and N$_{2}$H$^{+}$ lines, indicating that 
the abundances of NH$_{3}$ and N$_{2}$H$^{+}$ relative to 
carbon-chain molecules are very low, as observed in L1521E. 
We found that longer carbon-chain molecules 
such as HC$_{5}$N and C$_{4}$H are more abundant in TMC-1 than L1495B 
and L1521B, while those of sulfur-bearing molecules such as 
C$^{34}$S, CCS, and C$_{3}$S are comparable. 

These characteristic features of molecular abundances 
in L1495B and L1521B are similar to those of L1521E (Hirota et al. 2002). 
Therefore, L1495B and L1521B are also in the early stage of chemical 
evolution, and the depletion factor of heavy atoms are possibly lower 
than in other evolved cores such as L1544 (e.g. Bergin \& Langer 1997; 
Aikawa et al. 2003). 
The fact that L1495B, L1521B, and L1521E are found to be the cores with 
the lowest deuterium fractionation ratio of DNC/HNC and DCO$^{+}$/HCO$^{+}$ 
(Hirota et al. 2001; 2002; 2003) also supports this idea. 
These carbon-chain-producing regions, L1495B, L1521B, and L1521E, 
would be the best targets for observational studies on
initial conditions for dense core formation and star formation. 

\acknowledgements

We are grateful to Yuri Aikawa, Masatoshi Ohishi, Shuji Saito, 
and Norio Kaifu for valuable discussions. 
We are also grateful to the staff of Nobeyama Radio Observatory 
and Effelsberg 100 m telescope of 
MPIfR for their assistance in observations. 
We thank the anonymous referee for helpful comments and suggestions. 
TH thanks to the Inoue Foundation for Science 
(Research Aid of Inoue Foundation for Science) 
for the financial support. 
This study is partly supported by Grant-in-Aid from Ministry of 
Education, Science, Sports and Culture of Japan 
(14204013 and 15071201).

{}

\newpage

\newpage

\begin{deluxetable}{llrcccccccc}
\tabletypesize{\scriptsize}
\rotate
\tablenum{1}
\tablewidth{0pt}
\tablecaption{Observed lines 
\label{tab-observe}}
\tablehead{
\colhead{} & \colhead{} & 
  \colhead{$\nu$} &   \colhead{} & \colhead{$\mu$\tablenotemark{b}} & 
  \colhead{} & \colhead{$T_{B}$} &   $v_{lsr}$ (km s$^{-1}$) & 
  \colhead{$\Delta v$} &  \colhead{$T_{rms}$} & \\
\colhead{Source} & \colhead{Transition} & 
  \colhead{(MHz)} & \colhead{$S_{ul}$\tablenotemark{a}} & 
  \colhead{(Debye)} & 
  \colhead{Telescope} & \colhead{(K)} & \colhead{(km s$^{-1}$)} &
  \colhead{(km s$^{-1}$)} &  \colhead{(K)} & \colhead{Reference} 
}
\startdata
L1495B\tablenotemark{c}   
 & C$_{4}$H($N$=2-1,$J$=$\frac{5}{2}$-$\frac{3}{2}$,$F$=2-1) 
 & 19014.720 & 2.00 & 0.9 & MPIfR &  0.37  & 7.74 & 0.42 & 0.04  & \\ 
       & C$_{4}$H($N$=2-1,$J$=$\frac{5}{2}$-$\frac{3}{2}$,$F$=3-2) 
 & 19015.144 & 2.00 & 0.9 & MPIfR &  0.49  & 7.73 & 0.33 & 0.04  & \\
       & NH$_{3}$($J, K$ = 1,1) & 23694.506\tablenotemark{d}  & 1.50 & 1.47 & 
  MPIfR & 0.49  & 7.88 &  0.62 &  0.15  & \\
       & C$_{3}$S($J$=4-3)  & 23122.985  & 4.00 & 3.6 & 
  MPIfR & 1.29   &  7.59  &  0.24  &  0.11  & \\
       & HC$_{5}$N($J$=16-15)  & 42602.153  & 16.00 & 4.33 & 
  NRO & 0.36  & 7.65 &  0.47 & 0.04  & \\
       & CCS($J_{N}$=$4_{3}$-$3_{2}$)   & 45379.033  & 3.97 & 2.81 & 
  NRO & 2.1  & 7.69 &  0.31 &  0.09   & 1 \\
       & HC$_{3}$N($J$=5-4)   & 45490.316\tablenotemark{d}  & 5.00 & 3.72 & 
  NRO & 2.7  & 7.70 &  0.67 & 0.08  &  \\
       & H$^{13}$CO$^{+}$($J$=1-0)  & 86754.330  & 1.00 & 4.07 & 
  NRO & 0.88  & 7.75 &  0.46 &  0.10 & 2 \\
       & N$_{2}$H$^{+}$($J$=1-0) & 93173.777\tablenotemark{d} & 3.40 & 
   1.00 & NRO & \nodata & \nodata &  \nodata & 0.20  & \\
L1521B & C$_{4}$H($N$=2-1,$J$=$\frac{5}{2}$-$\frac{3}{2}$,$F$=3-2)  
 & 19014.720 & 2.00 & 0.9 & MPIfR &  0.37   &  6.44  &  0.21  &  0.10  & \\
       & C$_{4}$H($N$=2-1,$J$=$\frac{5}{2}$-$\frac{3}{2}$,$F$=2-1)  
 & 19015.144 & 2.00 & 0.9 & MPIfR & \nodata & \nodata & \nodata &  0.10  & \\
       & NH$_{3}$($J, K$ = 1,1) & 23694.506\tablenotemark{d}  & 1.50 & 1.47 & 
  MPIfR & 0.94  & 6.52 &  0.69 &  0.14  & \\
       & C$_{3}$S($J$=4-3) & 23122.985  & 4.00 & 3.6 & 
  MPIfR & 1.75   &  6.38  &  0.35  &  0.09  & \\
       & HC$_{5}$N($J$=16-15) & 42602.153  & 16.00 & 4.33 & 
  NRO & 0.87  & 6.44 &  0.43 & 0.06  & \\
       & CCS($J_{N}$=$4_{3}$-$3_{2}$)  & 45379.033  & 3.97 & 2.81 & 
  NRO & 2.8  & 6.2 &  0.40 &  0.1   & 3 \\
       & HC$_{3}$N($J$=5-4)  & 45490.316\tablenotemark{d}  & 5.00 & 3.72 & 
  NRO & 3.3  & 6.2 &  0.80 &  0.1  & 3 \\
       & H$^{13}$CO$^{+}$($J$=1-0)  & 86754.330  & 1.00 & 4.07 & 
  NRO & 1.14  & 6.57 &  0.56 &  0.10 & 2 \\
       & N$_{2}$H$^{+}$($J$=1-0) & 93173.777\tablenotemark{d} & 3.40 & 
   1.00 & NRO & \nodata & \nodata &  \nodata & 0.22  & \\
\enddata
\tablenotetext{ a}{ Intrinsic line strength}
\tablenotetext{ b}{ Dipole moment}
\tablenotetext{ c}{The position is (0",0"), except for 
C$_{3}$S and C$_{4}$H, which is at the (80\arcsec,40\arcsec) position. }
\tablenotetext{ d}{ Main hyperfine component}
\tablerefs{1: Hirota et al. (2001), corrected with the main beam efficiency 
of 0.7; 2: Hirota et al. (2001), corrected with the main beam efficiency 
of 0.5; 3: Suzuki et al. (1992), corrected with the main beam efficiency 
of 0.7}
\end{deluxetable}

\begin{deluxetable}{lcccccccc}
\tabletypesize{\scriptsize}
\rotate
\tablenum{2}
\tablewidth{0pt}
\tablecaption{Column densities of the selected molecules 
in unit of $10^{13}$ cm$^{-2}$
\label{tab-column}}
\tablehead{
\colhead{Molecule} & 
  \colhead{L1495B\tablenotemark{a}} &   \colhead{Reference} & 
  \colhead{L1521B} &   \colhead{Reference} &
  \colhead{L1521E} &   \colhead{Reference} & 
  \colhead{TMC-1} &  \colhead{Reference} 
}
\startdata
C$^{34}$S   & \nodata & \nodata & 0.56     & 2 & 1.25    & 1 & 0.73 & 1 \\
CCS              & 1.44     & 3 & 3.6      & 4 & 2.8     & 1 & 6.6  & 1 \\
C$_{3}$S         & 0.68     &   & 1.6      &   & 1.4     & 1 & 1.3  & 1 \\
C$_{4}$H         & 18.2     &   & 7.2      &   & 9.8     & 1 & 440  & 1 \\
HC$_{3}$N        & 2.1      &   & 4.1      & 4 & 2.3     & 1 & 17.1 & 1 \\
HC$_{5}$N        & 0.52     &   & 1.2      &   & 0.46    & 1 & 5.6  & 1 \\
NH$_{3}$         & 5.5      &   & 12.6     &   & 7.3     & 1 & 19   & 1 \\
N$_{2}$H$^{+}$   & $<$0.17  &   & $<$0.19  &   & $<$0.14 & 1 & 0.74 & 1 \\
H$^{13}$CO$^{+}$ & 0.046    & 3 & 0.063    & 3 & 0.106   & 1 & 0.14 & 1 \\
C$^{18}$O        & 190      & 5 & 300      & 5 & 170     & 1 & 330  & 1 \\
\enddata
\tablenotetext{ a}{The position is (0",0") except for 
C$_{3}$S and C$_{4}$H, which is at the (80\arcsec,40\arcsec) position. }
\tablerefs{1: Hirota et al. (2002) and references therein; 
  2: Hirota et al. (1998); 3: Hirota et al. (2001); 
  4: Suzuki et al. (1992); 5: Myers et al. (1983)}
\end{deluxetable}


\begin{thebibliography}{}
\bibitem[aik03]{aik03} Aikawa, Y., Ohashi, N., \& Herbst, E. 2003, ApJ, 593, 906
\bibitem[bei86]{bei86} Beichman, C. A., Myers, P. C., Emerson, J. P., 
  Harris, S., Mathieu, R., Benson, P. J., \& Jennings, R. E. 1986, ApJ, 
  307, 337
\bibitem[ben98]{ben98} Benson, P. J., Caselli, P., \& Myers, P. C. 1998, ApJ,
  506, 743
\bibitem[ben83]{ben83} Benson, P. J., \& Myers, P. C. 1983, ApJ, 270, 589
\bibitem[ben89]{ben89} Benson, P. J., \& Myers, P. C. 1989, ApJS,
  71, 89
\bibitem[ber97]{ber97} Bergin, E. A. \& Langer, W. D. 1997, ApJ, 486, 316
\bibitem[cas02]{cas02} Caselli, P., Benson, P. J., Myers, P. C., \& 
   Tafalla, M. 2002, ApJ, 572, 238
\bibitem[eli78]{eli78} Elias, J. H. 1978, ApJ, 224, 857
\bibitem[Eva01]{eva01} Evans, N. J., II, Rawlings, J. M. C., Shirley, Y., 
  \& Mundy, L. G. 2001, ApJ, 557, 193
\bibitem[]{} Fuente, A., Cernicharo, J., Barcia, A., \& G\'omez-Gonz\'alez, J. 
  1990, A\&A, 231, 151
\bibitem[gue82]{gue82} Gu\'elin, M., Friberg, P., \& Mezaoui, A. 
  1982, A\&A, 109, 23
\bibitem[hir92]{hir92} Hirahara, Y., et al. 1992, ApJ, 394, 539
\bibitem[hir98]{hir98} Hirota, T., Yamamoto, S., Mikami, H., \& 
  Ohishi, M. 1998, ApJ, 503, 717
\bibitem[hir01]{hir01} Hirota, T., Ikeda, M., \& Yamamoto, S. 2001, ApJ, 547, 814
\bibitem[hir03]{hir03} Hirota, T., Ikeda, M., \& Yamamoto, S. 2003, ApJ, 594, 859
\bibitem[hir02]{hir02} Hirota, T., Ito, T., \& Yamamoto, S. 2002, ApJ, 565, 359
\bibitem[kui96]{kui96} Kuiper, T. B. H., Langer, W. D., \& Velusamy, T. 1996, ApJ, 
  468, 761
\bibitem[lai00]{lai00} Lai, S. -P. \& Crutcher, R. M. 2000, ApJS, 128, 271
\bibitem[lee99]{lee99} Lee, C. W., Myers, P. C., \& Tafalla, M. 1999, 
  ApJ, 526, 788
\bibitem[lee03]{lee03}
  Lee, J.-E., Evans N. J., II, Shirley, Y. L., and Tatematsu, K. 
  2003, ApJ, 583, 789
\bibitem[mye83]{mye83} Myers, P. C., Linke, R. A., \& Benson, P. J. 1983, ApJ, 
   264, 517
\bibitem[oha99]{oha99} Ohashi, N., Lee, S. W., Wilner, D. J., 
  \& Hayashi, M. 1999, ApJ, 518, L41
\bibitem[oha00]{oha00} Ohashi, N. 2000, in IAU Symp. 197, Astrochemistry: 
  From Molecular Clouds to Planetary Systems, ed. Y. C. Minh \& E. F. van Dishoeck 
  (San Francisco: ASP), 61
\bibitem[ohi98]{ohi98} Ohishi, M. \& Kaifu, N. 1998, in Faraday Discussions 109, 
  Chemistry and Physics of Molecules and Grains in Space, 205
\bibitem[ola88]{ola88} Olano, C. A., Walmsley, C. M.,  \& Wilson, T. L. 1988, 
  A\&A, 196, 194
\bibitem[oni02]{oni02} Onishi, T., Mizuno, A., Kawamura, A., Tachihara, K., 
  \& Fukui, Y. 2002, ApJ, 575, 950
\bibitem[ott94]{ott94} Ott, M., Witzel, A., Quirrenbach, A., Krichbaum, T. P., 
  Standke, K. J., Schalinski, C. J., \& Hummel, C. A. 1994, A\&A, 284, 331
\bibitem[pra97]{pra97} Pratap, P., Dickens, J. E., Snell, R. L., Miralles, M. P., 
  Bergin, E. A., Irvine, W. M., \& Schloerb, F. P. 1997, ApJ, 486, 862
\bibitem[ruf97]{ruf97} Ruffle, D. P., Hartquist, T. W., Taylor, S. D., \& 
  Williams, D. A. 1997, MNRAS, 291, 235
\bibitem[sai00]{sai00} Saito, S., Ozeki, H., Ohishi, M., \& Yamamoto, S. 2000, ApJ, 535, 227
\bibitem[sai02]{sai02} Saito, S., Aikawa, Y., Herbst, E., Ohishi, M., Hirota, T., 
   Yamamoto, S., \& Kaifu, N. 2002, ApJ, 569, 836
\bibitem[suz92]{suz92} Suzuki, H., Yamamoto, S., Ohishi, M., Kaifu, N., Ishikawa, S., 
  Hirahara, Y., \& Takano, S. 1992, ApJ, 392, 551
\bibitem[taf04]{taf04} Tafalla, M. \& Santiago, J. 2004, A\&A, 414, L53
\bibitem[taf02]{taf02} Tafalla, M., Myers, P. C., Caselli, P., Walmsley, C. M., 
   \& Comito, C. 2002, ApJ, 569, 815
\bibitem[taf04b]{taf04b} Tafalla, M., Myers, P. C., Caselli, P., 
  \& Walmsley, C. M. 2004, A\&A, 416, 191
\bibitem[war99]{war99} Ward-Thompson, D., Motte, F., \& Andr\'e, P. 1999, 
  MNRAS, 305, 143
\end{thebibliography}
\end{document}